\documentclass[aps,prl,twocolumn,superscriptaddress]{revtex4}

\usepackage{graphicx}
\usepackage{bm}
\usepackage{amsmath}

\begin{document}

\title{Multi-fold Enhancement of Quantum Dot Luminescence in a Plasmonic Metamaterial}

\author{K. Tanaka}
\email{KenjiD.Tanaka@jp.sony.com} \affiliation{Optoelectronics
Research Centre, University of Southampton, SO17 1BJ, UK}
\affiliation{Sony Corporation, Shinagawa-ku, Tokyo, 141-0001, Japan}

\author{E. Plum}
\email{erp@orc.soton.ac.uk} \affiliation{Optoelectronics Research
Centre, University of Southampton, SO17 1BJ, UK}

\author{J. Y. Ou}
\affiliation{Optoelectronics Research Centre, University of
Southampton, SO17 1BJ, UK}

\author{T. Uchino}
\affiliation{School of Electronics and Computer Science, University
of Southampton, Southampton SO17 1BJ, UK}

\author{N. I. Zheludev}
\homepage{www.metamaterials.org.uk/} \email{niz@orc.soton.ac.uk}
\affiliation{Optoelectronics Research Centre, University of
Southampton, SO17 1BJ, UK}

\date{\today}

\begin{abstract}
We report that hybridizing semiconductor quantum dots with plasmonic
metamaterial leads to a multi-fold intensity increase and narrowing
of their photoluminescence spectrum. The luminescence enhancement is
a clear manifestation of the cavity quantum electrodynamics Purcell
effect that can be controlled by the metamaterial's design. This
observation is an essential step towards understanding loss
compensation in metamaterials with gain media and for developing
metamaterial-enhanced gain media.
\end{abstract}

\maketitle

Control of Joule losses is a key challenge for plasmonic and
metamaterial technologies. Losses hamper the development of negative
index media for super-resolution and optical cloaking devices, and
plasmonic data processing circuits. Lowering losses is also
crucially important for the performance of spectral filters, delay
lines and, in fact, practically any other metamaterial and plasmonic
applications \cite{Zheludev2010_Science}. Although using
superconducting metamaterials can largely eliminate losses in THz
and microwave metamaterials \cite{OE_2010_SuperconductingMM}, Joule
losses at optical frequencies are unavoidable. Recent works report
compensation of losses with gain in metamaterials aggregated with
semiconductor quantum dots (QDs) \cite{Eric2009_OE} and organic dyes
\cite{Xiao2010_Nature} embedded into the metal nanostructures.
Parametric metamaterials gain systems are also under investigation
in theory \cite{Wegener2008_OE, Fang2009_PRB,Sivan2010_OE}. Another
grand goal of active metamaterials research is to improve laser gain
media and to develop a `lasing spaser' device: a `flat' laser with
emission fueled by plasmonic excitations in an array of coherently
emitting meta-molecules \cite{Zheludev2008_spaser}. An essential
part of this development shall be the study of luminescence of
active material hybridized with plasmonic nanostructures that could
support collective, coherent plasmonic excitations in the lasing
spaser. Here we report the first study of photoluminescence of
semiconductor QDs hybridized with asymmetric split-ring plasmonic
metamaterial. This type of metamaterial supports a closed-mode
Fano-type excitation which has the key characteristics required for
the lasing spaser application: the mode is formed by collective
interactions between individual meta-molecules that shall ensure
coherent laser action \cite{Fedotov2007_PRL_TrappedModes}. In this
letter, we experimentally demonstrate that the photoluminescence
properties of QDs can be greatly enhanced by the plasmonic
metamaterial.

\begin{figure}[tb!]
\includegraphics[width=85mm]{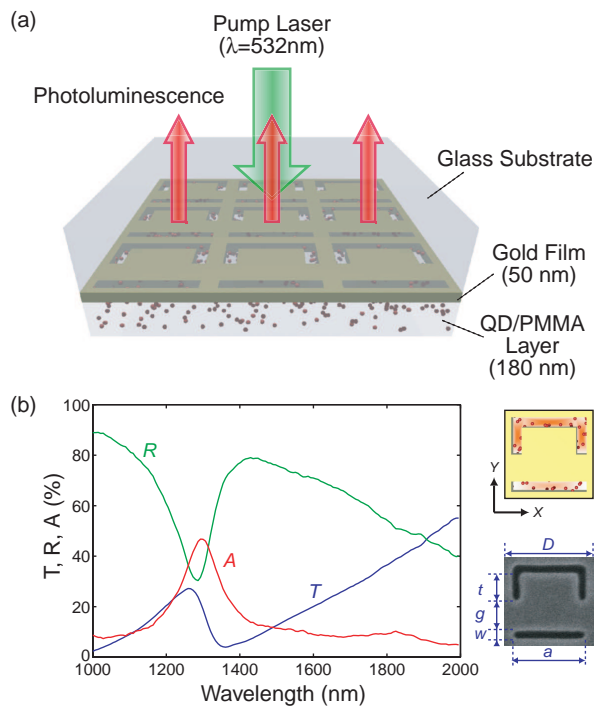}
\caption{\label{fig1}(color online). (a)~Schematic of a plasmonic
metamaterial functionalized with quantum dots (QDs). (b)~Measured
spectra of transmission $T$, reflection $R$, and absorption $A$ for
a QD-coated metamaterial with a unit cell size of $D=545$~nm. The
incident wave is polarized parallel to the $y$-axis. The insets show
a sketch of QDs in the resonant mode volume as seen from the
substrate side and a scanning electron micrograph of the unit cell
(without QDs). Feature sizes: unit cell $D=545$~nm, horizontal slit
$a=470$~nm, top vertical slit and gap $t=g=170$~nm and slit width
$w=65$~nm.}
\end{figure}

Figure~\ref{fig1}(a) schematically illustrates a plasmonic
metamaterial combined with QDs. The metamaterials studied here
consist of periodic arrays of asymmetrically split ring slits
(negative structure), which have been successfully applied to
switching, nonlinear and sensor applications
\cite{NatMat_2010_FanoResonance}. The metamaterial arrays with a
total size of $40\times 40~\mu\text{m}$ each were fabricated by
focused ion beam milling in a 50nm-thick gold film on a glass
substrate [see inset of Fig.~\ref{fig1}(b)].

In order to systematically investigate the correlation between QD
photoluminescence spectrum and the spectral position of the Fano
plasmonic metamaterial resonance, we manufactured five metamaterial
arrays with different unit cell sizes ranging from $D=545$~nm to
645~nm, slit width $w=65$~nm and a fixed ratio of $t/g=1$. We used
lead sulfide (PbS) semiconductor quantum dots from Evident
Technologies with a luminescence peak around 1300~nm and mean core
diameter of 4.6~nm. These QDs were dispersed in
polymethylmethacrylate (PMMA) and the QD/PMMA solution was then
spin-coated onto the metamaterial arrays forming a 180~nm thick
layer. We estimate the QD area density on the array to be
$1.6\times10^5~\mu\text{m}^{-2}$ and thus approximately $4000$
quantum dots per meta-molecule are trapped in the groves of the
structure. Spectra (transmission, reflection, and absorption) and
photoluminescence characteristics of the metamaterials with QDs were
measured using a microspectrophotometer. In the photoluminescence
measurements, the QDs were optically pumped by a frequency doubled
CW~YAG laser ($\lambda=532$~nm) from the substrate side through the
metamaterial array [see Fig.~\ref{fig1}(a)]. The YAG laser was
focused to a ($\sim100~\mu\text{m}$) spot with intensity
35~W/$\text{cm}^2$ by the microscope objective (N.A.$=0.28$).
Photoluminescence emitted from the QDs was collected by the same
objective with a polarizer so that only a selected polarization
component of photoluminescence could be detected. Fig.~\ref{fig1}(b)
shows the spectral characteristics of the QD/PMMA-coated
metamaterial with $D=545$~nm in the absence of the pump. The
absorption spectrum shows a narrow resonance peak around 1300~nm
($Q\text{-factor}\simeq11$).


\begin{figure}[tb!]
\includegraphics[width=75mm]{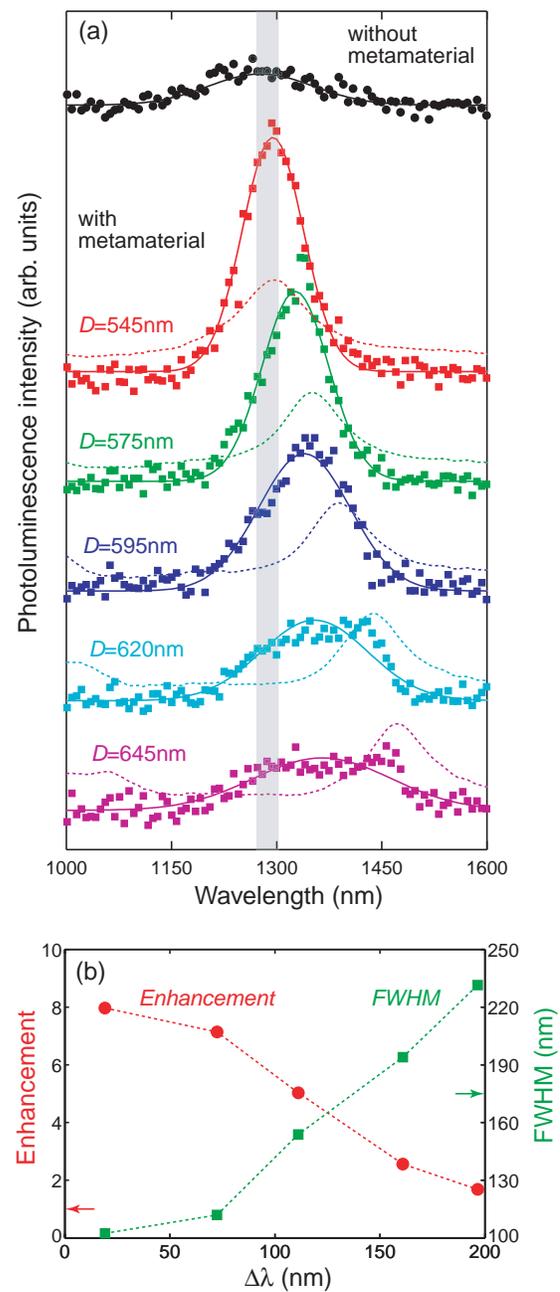}
\caption{\label{fig2}(color online). Photoluminescence controlled by
plasmonic metamaterials. (a)~Photoluminescence spectra of the QDs
without (top) and with metamaterial layer. A best-fit Gaussian curve
is plotted with each spectrum (solid line). The dotted line
indicates the metamaterial absorption spectrum. (b)~Intensity
enhancement and full width at half maximum (FWHM) of the
QD-metamaterial photoluminescence spectra. Enhancement is normalized
to the photoluminescence peak intensity measured without a
metamaterial layer. For comparison, reference values measured
without a metamaterial layer are indicated by arrows.}
\end{figure}

The photoluminescence of the metamaterial functionalized with QDs is
presented in Fig.~\ref{fig2}. Here we measure the $y$-polarization,
for which the metamaterial's Fano-mode can be excited, see
Fig.~\ref{fig1}(b). The photoluminescence spectrum of the QD/PMMA
layer on the glass substrate (i.e.~without metamaterial) is shown at
the top of Fig.~\ref{fig2}(a) and peaks at $\lambda_0=1280$~nm which
is indicated by the shaded area.

The presence of the plasmonic metamaterial drastically changes the
QD photoluminescence characteristics: it leads to a multi-fold
intensity enhancement as well as spectral narrowing of the
photoluminescence peak. For instance, for $D=545$~nm the
photoluminescence peak intensity is enhanced by a factor of 8, while
the full width at half maximum (FWHM) of the photoluminescence peak
is decreased to approximately 100~nm compared to 176~nm without
metamaterial. Here the metamaterial's absorption resonance
wavelength $\lambda_{\text{abs}}$ almost perfectly matches the QD
emission wavelength $\lambda_0$ (see red dashed absorption
spectrum). This suggests that the drastic photoluminescence
enhancement results from interaction between the excited-state gain
medium (QDs) and the surface plasmon resonance
\cite{Lakowicz2001,Aslan2005,Bergman2003_spaser,Stockman2008_spaser},
and, in particular, can be understood in terms of the cavity quantum
electrodynamics (QED) Purcell effect
\cite{Vahala2004_microcavities,Book_2006_Nano-Optics} as discussed
later. Such coupling between QD-excitons and metamaterial surface
plasmons must be sensitive to a mismatch
$\Delta\lambda=\lambda_{\text{abs}}-\lambda_0$. Indeed, when the
metamaterial resonance is red-shifted, by increasing the unit cell
size from $D=545$~nm to 645~nm, the photoluminescence spectrum is
weakened, broadened and distorted. In all cases the
photoluminescence peak is shifted from its original position
$\lambda_0$ towards the respective metamaterial's absorption
resonance $\lambda_{\text{abs}}$. For a large mismatch
$\Delta\lambda>150$~nm (i.e.~$D\geq620$~nm), the photoluminescence
spectrum becomes non-Gaussian and appears to develop two peaks close
to $\lambda_0$ and $\lambda_{\text{abs}}$, respectively.
Intriguingly, the observed photoluminescence red-shift becomes quite
large, reaching almost 200~nm. The intensity enhancement and FWHM of
the photoluminescence spectra are summarized in Fig.~\ref{fig2}(b)
as a function of the mismatch $\Delta\lambda$. Narrow
photoluminescence spectra with greatly enhanced intensity are
observed when the QD-luminescence matches the metamaterial resonance
wavelength (i.e. small $\Delta\lambda$). On the other hand, for a
large mismatch ($\Delta\lambda>150$~nm) the photoluminescence
spectrum becomes even broader than it is without the metamaterial
layer. Here we note that all metamaterial samples had almost
identical transmission levels at the pump wavelength (532~nm), and
therefore there was no significant difference in pump power reaching
the active layer.

One might attempt to explain such broadening as a result of
filtering the QD luminescence spectrum through the metamaterial,
thus assuming no plasmon-exciton interaction. However, such a simple
explanation can be ruled out, as it does not explain any
photoluminescence enhancement resulting from the presence of the
metamaterial. Furthermore, the convoluted spectrum of QD
luminescence and metamaterial transmission does not agree well with
the photoluminescence measurements.

\begin{figure}[tb!]
\includegraphics[scale=1.0]{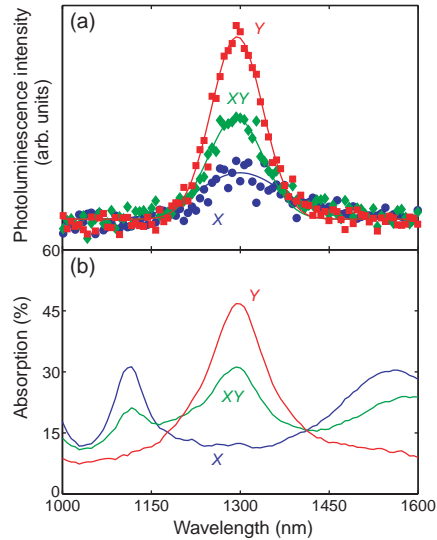}
\caption{\label{fig3}(color online). Polarization dependence of (a)
photoluminescence and (b) absorption, measured for the metamaterial
with $D=545$~nm. Polarizations $x$ and $y$ are introduced in
Fig.~\ref{fig1}, while $xy$ is the intermediate polarization.}
\end{figure}

The metamaterials studied here have profound polarization-dependent
properties. While a strong plasmonic Fano-resonance is excited by
$y$-polarization [see Fig.~\ref{fig1}(b)], this resonance vanishes
for the orthogonal polarization \cite{Eric2009_OE}. This
polarization dependence may be expected to affect the interaction
with the isotropic QDs, and hence the polarization dependence of the
photoluminescence was measured as illustrated by Fig.~\ref{fig3} for
the metamaterial with $D=545$~nm. By changing the polarization state
from $y$ to $x$, the absorption spectrum becomes featureless around
$\lambda_0$ [Fig.~\ref{fig3}(b)]. The corresponding
photoluminescence drastically degrades [Fig.~\ref{fig3}(a)],
providing additional evidence that the photoluminescence spectrum is
controlled by the plasmonic resonance.

We argue that the observed enhancement of photoluminescence can be
understood in terms of the cavity QED Purcell effect
\cite{Vahala2004_microcavities,Book_2006_Nano-Optics}. Indeed, the
spontaneous emission decay rate is proportional to the density of
photon states that the photonic environment offers for spontaneous
decay. Thus the internal dynamics of a quantum system are controlled
by a photonic environment that is resonant with radiative
transitions of the source. Enhancement of the radiation rates has
been seen in various systems including QDs in nanocavities and
photonic crystals. In our experiments the ensemble of QDs with its
exciton emission line is placed at a resonant plasmonic
metamaterial. The metamaterial creates an environment equivalent to
a microcavity with a quality factor $Q$, and a mode confined in an
ultrasmall volume $V$ that enhances the density of photon states
leading to the Purcell factor enhancement of luminescence:
$F_p=\tfrac{3}{4\pi^2}\left(\tfrac{\lambda}{n}\right)^3\tfrac{Q}{V}$.
Here $n$ is the refractive index of the medium and $\lambda$ is the
wavelength.

For the sake of rough estimate, the mode volume is calculated by
$V=2(a+t)wh$ (where $a=470$~nm, $t=170$~nm, $w=65$~nm, and
$h=50$~nm), i.e. the mode is assumed to be confined in the slits of
the metamaterial metal film of thickness $h$. With
$\lambda=1300$~nm, $Q=11$, and $n=1.48$, this gives the following
value for the Purcell factor $F_p=136$. This is of the same order of
magnitude as the experimentally observed enhancement of overall
luminescence $\mu=8$, corrected for the fraction of QDs in the slits
$f=V/(D^2p)$, where $p=180$~nm is the thickness of the QD/PMMA
layer: $F_{p,exp}=\mu/f=103$. Here we note that the general Purcell
enhancement formula used in our calculations only gives approximate
values for luminescence enhancement in plasmonic systems
\cite{Koenderink_2010_NanoPurcellFactor}.

\begin{figure}[t!]
\includegraphics[scale=1.0]{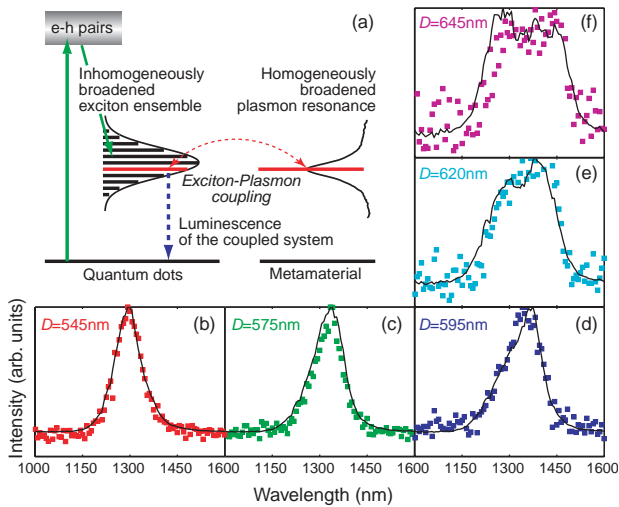}
\caption{\label{fig4}(color online). Nature of photoluminescence
change in the plasmonic metamaterial. (a)~Energy diagram of the
QD-metamaterial coupled system. (b)-(f) Comparison of the measured
photoluminescence (data points) with
$\chi_A(\lambda)=PL_0(\lambda)\times A(\lambda)$ (lines) for
metamaterials with different unit cell sizes ranging from $D=545$~nm
to 645~nm. Here, $PL_0(\lambda)$ is normal QD photoluminescence
without metamaterial structure, and $A(\lambda)$ is the
metamaterial's absorption spectrum.}
\end{figure}

Peculiarity of our experimental conditions in comparison with
numerous reports on the Purcell factor enhancement of luminescence
of individual QDs is in a large number of QDs located within the
mode volume ($\sim4000$): the exciton line is inhomogeneously
broadened due to a natural variation of the QD sizes
[Fig.~\ref{fig4}(a)]. Here, detuning of the plasmon resonance from
the centre of the exciton emission line leads to the Purcell
enhancement being applied to the wing of the emission line as it is
clearly manifested by the transformation of the photoluminescence
spectrum presented in Fig.~\ref{fig2}. The photoluminescence spectra
resulting from this Purcell enhancement may be expected to be
proportional to $\chi_A(\lambda)=PL_0(\lambda)\times A(\lambda)$,
where $PL_0(\lambda)$ is the normal QD photoluminescence spectrum
without metamaterial structure and $A(\lambda)$ is the absorption
spectrum of the metamaterial array (a measure of the local density
of states). As illustrated by Fig.~\ref{fig4}(b)-(f),
$\chi_A(\lambda)$ (lines) is in excellent agreement with the
measured photoluminescence (data points) in all cases, providing
further evidence for the Purcell effect and plasmon-exciton coupling
in the QD-metamaterial system.

We argue that in a coupled QD-plasmonic metamaterial system the
resonant enhancement of luminescence can be exploited for increasing
optical gain and thus for the development of compact, low-threshold
lasing devices. At the same time it is not clear yet what effect the
profound Purcell enhancement of luminescence has on the
metamaterial's Joule losses as it reduces the fraction of energy
that is transferred to the plasmonic system.

In summary, we have experimentally demonstrated multi-fold
enhancement and substantial spectral narrowing of photoluminescence
from semiconductor quantum dots resulting from resonant coupling to
a plasmonic metamaterial. We have shown that the intensity
enhancement and spectral width of the photoluminescence in the
combined system are controlled by the spectral overlap of the
emission peak of free QDs and the metamaterial's plasmonic
resonance, and thus this effect is linked to exciton-plasmon
coupling between QDs and metamaterial. The observed
photoluminescence enhancement provides the first and clear
demonstration of the cavity QED Purcell effect in metamaterials.

\begin{acknowledgments}
Financial support of the Engineering and Physical Sciences Research
Council, UK is acknowledged.
\end{acknowledgments}


\end{document}